\documentclass{elsart}

 \usepackage{graphicx}

\usepackage{amssymb,amsmath}
\usepackage{amssymb,pxfonts,txfonts}
\usepackage{bm}
\usepackage{dcolumn}

\begin{document}

\begin{frontmatter}



\title{First-order phase transition of triangulated  surfaces on a spherical core}


\author{Hiroshi Koibuchi}
\ead{koibuchi@mech.ibaraki-ct.ac.jp}

\address{Department of Mechanical and Systems Engineering, Ibaraki National College of Technology, 
Nakane 866, Hitachinaka,  Ibaraki 312-8508, Japan}

\begin{abstract}
We study an intrinsic curvature model defined on fixed-connectivity triangulated lattices enclosing a spherical core by using the canonical Monte Carlo simulation technique. We find that the model undergoes a discontinuous transition of shape transformation between the smooth state and a collapsed state even when the core radius $R$ is sufficiently large; the transition depends on $R$. The origin of the multitude of transitions is considered to be a degeneracy of the collapsed states.  We also find that the Gaussian bond potential $S_1/N$, which is the sum of bond length squares, discontinuously changes at the transition. The discontinuity in $S_1/N$ implies a possibility of large fluctuations of the distance between lipids, or the density of lipids, in biological membranes such as giant vesicles or liposomes enclosing some materials. 

\end{abstract}

\end{frontmatter}

\section{Introduction}\label{intro}
A lot of attention has been paid to the phase structure of surface models for strings and membranes \cite{P-L-1985PRL,Paczuski-Kardar-Nelson-PRL1988,DavidGuitter-1988EPL,BORELLI-KLEINERT-SCHAKE-PLA2000,Kownacki-Mouhanna-PRE2009}. The phase transitions of the models are closely connected to a variety of morphology in membranes \cite{NELSON-SMMS2004,Gompper-Schick-PTC-1994,Bowick-PREP2001,WIESE-PTCP19-2000,SEIFERT-LECTURE2004,WHEATER-JP1994}. The Hamiltonian is defined by the squared mean curvature or the extrinsic curvature on two-dimensional smooth surfaces embedded in ${\bf R}^3$ \cite{HELFRICH-1973,POLYAKOV-NPB1986}, and the statistical mechanical models are obtained by discretizing the Hamiltonian on the triangulated surfaces \cite{KANTOR-NELSON-PRA1987,GOMMPER-KROLL-SCIENCE1992,GOMMPER-KROLL-PRA1992-PRE1995,GOMPPER-KROLL-SMMS2004,AMBJORN-NPB1993}. In numerical studies, phantom surfaces are always assumed as membranes for computational efficiency; a surface which is allowed to self-intersect is called the phantom surface. The phase structure of the phantom surfaces has been extensively studied and is partly clarified such that the spherical surface models undergo a first-order transition of shape fluctuations \cite{Muenkel-Heermann-JPhys-1993,Muenkel-Heermann-PRL-1995,KD-PRE2002,NISHIYAMA-PRE-2004,KOIB-PRE-2004,KOIB-PRE-2005,KOIB-NPB-2006}.

Membranes are always understood as boundary surfaces separating interior materials from exterior environment. Membranes can shrink/swell if the interior material is liquid. Liquid can leak out/in through pores or by some other reasons, and consequently the membrane fluctuates as a two-dimensional surface. Thus, enclosed materials, such as liquid, and their property are considered to be included in the surface model as the bending rigidity. To the contrary, hard materials can not simply be included in the surface model. Surface fluctuations of membranes are considerably suppressed by such hard materials. Membranes which enclose the hard materials never shrink. 

Thus, the phase structure of membranes enclosing hard materials is expected to be different from the phase structure of membranes without the hard materials. It is possible that the collapse phenomenon does not occur on the surfaces enclosing a spherical core. The surface is prohibited from changing the orientation upside down owing to the core inside the surface even when the surface is phantom.

Therefore, it is interesting to study the phase structure of surface models with a spherical core and to see the dependence of the phase structure on the core radius $R$. In this paper, we study a fixed-connectivity spherical surface model in \cite{KOIB-EPJB-2004} with a spherical core by using the canonical Monte Carlo (MC) simulation technique. The model is an intrinsic curvature model \cite{Baillie-Johnston-PRD-1993-1994,BEJ,BIJJ} and is different from the extrinsic curvature ones in \cite{Muenkel-Heermann-JPhys-1993,Muenkel-Heermann-PRL-1995,KD-PRE2002,NISHIYAMA-PRE-2004,KOIB-PRE-2004,KOIB-PRE-2005,KOIB-NPB-2006}. The intrinsic curvature is independent of the surface shape, and hence its value is expected to be almost independent of whether the surface encloses the spherical core or not. 
\section{Model}\label{model}
\subsection{Fixed-connectivity phantom surface models}\label{surface_models}
In this subsection, we make a brief survey of the triangulated surface models and their phase structures. 

The Hamiltonian $S$ is given by a linear combination of the bond potential $S_1$ and the curvature energy $S_2$ such that $S\!=\!S_1\!+\!bS_2$, where $b[kT]$ is the bending rigidity. The continuous Hamiltonian is discretized on triangulated surfaces of sphere topology in ${\bf R}^3$ for numerical studies.  Muenkel and Heermann reported in Ref. {\cite{Muenkel-Heermann-JPhys-1993}} that the model undergoes a phase transition between the smooth phase and the crumpled phase on both of the sphere and a torus in ${\bf R}^3$, where the Hamiltonian {$S$} includes an intrinsic curvature term, and moreover that a self-avoiding surface model undergoes a discontinuous transition {\cite{Muenkel-Heermann-PRL-1995}}. The role of $S_1$ is to make the mean bond length constant. Therefore, a Lennard-Jones type potential can be assumed as $S_1$ \cite{Muenkel-Heermann-PRL-1995,KD-PRE2002} and, a hard wall potential is also possible \cite{KOIB-NPB-2006}. In Ref.\cite{KANTOR-NELSON-PRA1987} of Kantor and Nelson, a hard wall and hard core potential is assumed as $S_1$; the model is well known as the ball-spring model. Gompper and Kroll extensively studied the surface model including the ball-spring model on both phantom and self-avoiding surfaces \cite{GOMMPER-KROLL-SCIENCE1992,GOMMPER-KROLL-PRA1992-PRE1995,GOMPPER-KROLL-SMMS2004}.  On the other hand, the role of $S_2$ is to make the surface smooth at sufficiently large finite $b$. 

In the model of Helfrich and Polyakov, $S_1$ is given by the Gaussian bond potential; $S_1\!=\!\sum_{ij}(X_i\!-\!X_j)^2$, where $X_i(\in {\bf R}^3)$ is the vertex position, and $S_2\!=\!\sum_{ij}(1\!-\!{\bf n}_i\cdot {\bf n}_j)$, where ${\bf n}_i$ is a unit normal vector of the triangle $i$. We call this model as the {\it canonical} surface model. The canonical model, as mentioned in the Introduction, was reported to have a first-order phase transition separating the collapsed phase from the smooth phase \cite{KOIB-PRE-2005}. Note also that the canonical surface model is expected to have a continuous transition from theoretical calculations on the basis of the renormalization group \cite{DavidGuitter-1988EPL,BORELLI-KLEINERT-SCHAKE-PLA2000,Kownacki-Mouhanna-PRE2009,Essafi-Kownacki-Mouhanna-PRL2011}. 

The intrinsic curvature model is obtained by replacing $S_2$ with the intrinsic curvature energy $S_2^{\rm int}\!=\!-\sum_{i} \log \left(\delta_i/2\pi\right)$, where $\delta_i$ is the sum over angles of triangles meeting at the vertex $i$. This term {$-\sum_{i} \log \delta_i$} corresponds to a weight {$\Pi_i q_i^\sigma=\exp\left( \sigma \sum_i\log q_i\right)$} for the integration of the partition function, where {$q_i$} is the coordination number of the vertex {$i$} and {$\sigma\!=\!d/2\!=\!3/2$} {\cite{DAVID-NPB-1985}}. As mentioned above, surface models defined by a Hamiltonian which includes the term {$-(3/2)\sum_i \log q_i $} are extensively studied {\cite{Muenkel-Heermann-JPhys-1993,Baillie-Johnston-PRD-1993-1994,BEJ,BIJJ}}, while the constant parameter $\sigma$ is extended to a variable one {$\lambda$}, and a model with such an intrinsic curvature term {$\lambda\sum_i|q_i\!-\!6|$} was also extensively studied {\cite{Baillie-Johnston-PRD-1993-1994}}. $S_2^{\rm int}\!=\!\sum_{i}|\delta_i\!-\!2\pi|$ or $S_2^{\rm int}\!=\!\sum_{i}(\delta_i\!-\!2\pi)^2$ can also be assumed as the intrinsic curvature energy, where the deficit angle $\delta_i\!-\!2\pi$ is exactly zero on the planar lattices. It is apparent that $\delta_i\!=\!2\pi$ on the cylindrical lattices if the triangulation is sufficiently fine. Thus, $S_2^{\rm int}$ does not always depend on the surface shape in contrast to the extrinsic curvature energy. However, numerical studies indicate that the intrinsic curvature model undergoes a first-order transition of shape transformation; not only the collapsed phase but also the smooth phase is stable \cite{KOIB-EPJB-2004}. This is a non-trivial result in the sense that $S_2^{\rm int}$ uniquely determines the surface shape even though $\delta_i$ is not always dependent on the surface shape. 

The phase transition of the intrinsic curvature model can be seen on spheres, tori, and disks \cite{KOIB-EPJB-2004,KOIB-PLA-2006-1,KOIB-PLA-2005}. Hence the transition is independent of these three types of topology.  

As mentioned above, $S_2^{\rm int}$ is zero not only on the planar surface but also on the tubular surface. Thus, the smooth phase is considered to be degenerate in the model on the disk surface, because the planar disk can turn a tubular surface without a cost of $S_2^{\rm int}$. However, the smooth surface is almost planar on the disk surface \cite{KOIB-PLA-2005}. This implies that the smooth phase is "almost degenerate". This allows us to consider that the collapsed phase is also degenerate or almost degenerate. Many of the collapsed surfaces are expected to belong to a single value of $S_2^{\rm int}$ or to a narrow and continuous range of $S_2^{\rm int}$. We numerically check this expectation in this paper and show that it is correct. 

\subsection{Model with a hard-core potential}\label{model_in_this_paper}
\begin{figure}[ht!]
\centering
\includegraphics[width=10cm]{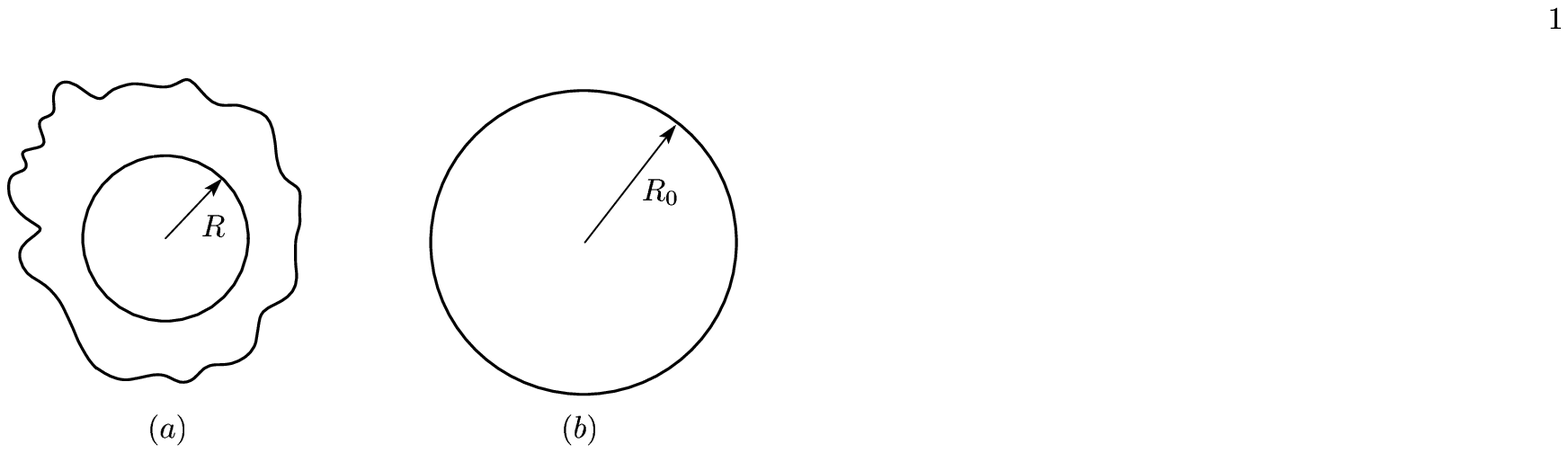}
\caption{ (a) A schematic view of a surface enclosing a spherical core of radius $R$, (b) the smooth surface configuration of triangulated sphere of radius $R_0(N)$, which satisfies $S_1/N\!=\!3/2$. The surface in (b) is expected to appear in the limit of $\alpha\!\to\!\infty$. 
 } 
\label{fig-1}
\end{figure}
In this subsection, we define an intrinsic curvature model with a hard core potential. Figure \ref{fig-1}(a) shows a spherical surface enclosing a spherical core of radius $R$. The surface size is always greater than the core size even in the collapsed phase. 

The triangulated spherical lattice in this paper is identical to that in \cite{KOIB-PRE-2005} and is constructed from the icosahedron by splitting the edges (faces) into $\ell$ ($\ell^2$) pieces. The total number of vertices is thus given by $N\!=\!10\ell^2\!+\!2$, the total number of bonds $N_B$ and the total number of triangles $N_T$ are given by $N_B\!=\!30\ell^2$ and $N_T\!=\!20\ell^2$. The coordination number $q$ is $q\!=\!6$ almost everywhere on the lattice excluding the $12$ vertices, which are the vertices of the icosahedron and of $q\!=\!5$.  

The partition function $Z$ is defined by
\begin{equation} 
\label{Part-Func}
 Z = \int \prod _{i=1}^{N} d X_i \exp\left[-S(X)\right],
\end{equation} 
where $\int\prod _{i=1}^{N} d X_i$ is the $3N$-dimensional integrations, and $X_i\left(\in {\bf R}^3\right)$ is the vertex position $i$. The Hamiltonian $S(X)$ denotes that $S$ depends on the surface shape $X(=\{X_1,X_2,\cdots,X_N\})$. The Hamiltonian is defined by a linear combination of the Gaussian bond potential $S_1$, the deficit angle energy $S_2^{\rm int}$, and the hard-core potential $U(R)$ such that
\begin{eqnarray}
\label{Disc-Eneg_1}
&& S(X)=S_1 + \alpha S_2^{\rm int} + U(R),  \nonumber \\
&& S_1=\sum_{(ij)} \left(X_i-X_j\right)^2,\quad S_2^{\rm int}=-\sum_{i} \log \left(\delta_i/2\pi\right),\\
&& U(R)=\sum_i U_i(R). \nonumber
\end{eqnarray} 
 $\sum_{(ij)}$ in $S_1$ denotes the sum over bonds $(ij)$. The symbol $\alpha[kT]$ is the curvature coefficient, and $\delta_i$ in $S_2^{\rm int}$ is the sum over angles of triangles meeting at the vertex $i$;  $\delta_i$ was introduced in the previous subsection \ref{surface_models}. $\sum_i$ in $S_2^{\rm int}$ denotes the sum over vertices $i$, and $U_i(R)$ in $U(R)$ is given by
\begin{equation}
\label{wall-potential} 
U_i(R)= \left\{
       \begin{array}{@{\,}ll}
        0 & \; \left(|X_i|>R \right), \\
      \infty & \; \left({\rm otherwise}\right). 
       \end{array} 
       \right. \\ 
\end{equation} 
We should note that the center of the core is fixed at the origin of ${\bf R}^3$, while the center of mass of the surface is not fixed in the integrations of $Z$ in contrast to the integrations of $Z$ of the models without the core. The translational zero mode of the surface can influence the results if the surface configurations are partly dominated by such configurations that are identical to each other up to translation. However, such configurations are not expected, moreover, the core protects the surface from the translation. The collapsed surface is at least prohibited from the translation, because the size of collapsed surface is not so larger than the core size. As for the smooth surface, the surface translation is also expected to be almost suppressed.

We should note that the scale invariance of the partition function is broken by the term {$U(R)$} {\cite{WHEATER-JP1994}}. As a consequence, the relation $\langle S_1/N \rangle \!=\!3/2$, which holds in the model without the cores, is expected to be violated because of the term $U(R)$ and the scale invariance of the partition function \cite{WHEATER-JP1994}. 

\begin{table}[hbt]
\caption{ The assumed values of $N$, $R_0$, and $R$. The core size $R$ ranges from $R(50\%)$ to $R(20\%)$. }
\label{table-1}
\begin{center}
 \begin{tabular}{ccccc}
 $N$ &  $R_0$ & $R (50\%)$ & $R (30\%)$& $R (20\%)$  \\
 \hline
  1442  & 7.046  & 3.523  & 2.1138 & 1.4092  \\
  2562  & 9.39   & 4.695  & 2.817  & 1.878   \\
  4842  & 12.9   & 6.45   & 3.87   & 2.58    \\
  6252  & 14.66  & 7.33   &   -    &   -     \\
  8412  & 17.01  & 8.505  &   -    &   -     \\
  10242 & 18.76  & 9.38   &   -    &   -     \\
  12252 & 20.52  & 10.26  &   -    &   -     \\
 \hline
 \end{tabular} 
\end{center}
\end{table}
We assume several different values of $R$ such that $R/R_0\!=\!50\%$, $R/R_0\!=\!30\%$, $R/R_0\!=\!20\%$,  and $R/R_0\!=\!0\%$. The symbol $R_0$ denotes the radius of smooth sphere, where the bond length squares $(X_i\!-\!X_j)^2$ satisfies $\sum_{(ij)}(X_i\!-\!X_j)^2/N(\!=\!S_1/N)\!=\!1.5$. $R_0$ is identical with the radius of the equilibrium surface in the limit of $\alpha\!\to\!\infty$ and is determined only by $N$; $R_0$ is a parameter that depends only on $N$. Figure \ref{fig-1}(b) shows a sphere section of radius $R_0$. The core radius $R$ is also determined by both $N$ and the ratio $R/R_0$. The assumed values of $R_0$ and $R$ are shown in Table \ref{table-1}. Note also that {$R_0^2/N$}, and hence {$R^2/N$} also, becomes a constant independent of {$N$} including the limit of {$N\!\to\!\infty$}. In fact, the area {$4\pi R_0^2$} of the surface of radius {$R_0$} is proportional to $N$, because the mean bond length squares is a constant, which is independent of $N$ due to the fact that {$S_1/N(=1.5)$}.

\section{Monte Carlo technique}\label{MC-Techniques}
The vertex position $X$ on the triangulated surface is sequentially updated by using the canonical Metropolis MC technique. A new position $X^\prime\!=\!X\!+\!\delta X$ is accepted with the probability ${\rm Min}[1,\exp (-\delta S)]$, where $\delta S\!=\!S({\rm new})\!-\!S({\rm old})$. The small variation $\delta X$ is randomly generated in a sphere of a given radius, which is fixed to maintain approximately $50\%$ acceptance rate of the Metropolis accept/reject. The Metropolis algorithm is applied only to vertices, of which $X^\prime$ is located outside the core. New positions $X^\prime$ inside the core are rejected. This acceptance rate of $X^\prime$ is almost independent of $R$ and, almost all $X^\prime$ are accepted. Thus, the total acceptance rate depends almost only on the Metropolis acceptance rate. We should note that a small core, such as the core of the {$N\!=\!1442$} surface under {$R/R_0\!=\!10\%$}, plays no role for the potential {$U(R)$}. In fact, a new position {$X^\prime$} on such small surfaces can move from {$X$}, which is outside of the core, to opposite side of the core. To check that a core is effective as {$U(R)$}, it is helpful to see whether {$S_1/N\!=\!1.5$} holds or not at the transition, because the relation {$S_1/N\!=\!1.5$} is violated only in the collapsed phase due to the potential {$U(R)$} as mentioned above. 

The total number of MC sweeps (MCS) after sufficiently large number of thermalization MCS is approximately $5\times10^8\sim 7\times10^8$ at the transition region of the surfaces of $N\!=\!8412$, {$N\!=\!10242$}, and $N\!=\!12252$, and the total number of MCS is relatively small at the non-transition region and on the smaller sized surfaces.

\section{Results}\label{Results}
\begin{figure}[ht!]
\centering
\includegraphics[width=10cm]{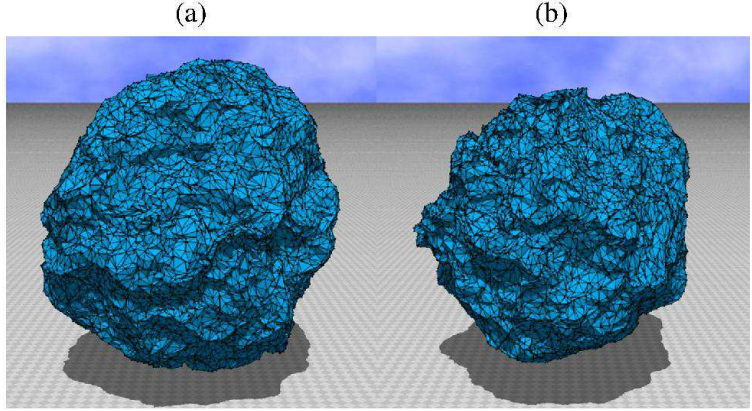}
\caption{Snapshots of the surface of size $N\!=\!12252$ at (a) $\alpha\!=\!760$ in the smooth phase and at (b) $\alpha\!=\!730$ in the collapsed phase. The ratio of $R$ to $R_0$ is $R/R_0\!=\!50\%$. The mean square size is about (a) $X^2\!=\!175$ and (b) $X^2\!=\!145$, which correspond to the mean radius $\sqrt{175}\simeq 13.2$ and $\sqrt{145}\simeq 12$. The snapshots are drawn in the same scale.} 
\label{fig-2}
\end{figure}
Snapshots of surfaces of $N\!=\!12252$ are shown in Figs. \ref{fig-2}(a) and \ref{fig-2}(b), which are separated by a first-order transition. Both surfaces appear to be in the smooth phase. However, the surfaces are different from each other because of the transition, which is the main result of this paper. The surfaces in Figs. \ref{fig-2}(a) and \ref{fig-2}(b) are respectively obtained at $\alpha\!=\!760$ and $\alpha\!=\!730$ under the condition $R/R_0\!=\!50\%$. The mean square size $X^2$ is $X^2\!=\!175$  and $X^2\!=\!145$ on the surfaces in Figs. \ref{fig-2}(a) and \ref{fig-2}(b), respectively, where $X^2$ is defined by 
\begin{equation}
\label{X2}
X^2={1\over N} \sum_i \left( X_i-\bar X\right)^2, \quad \bar X={1\over N} \sum_i X_i.
\end{equation}
We should note that the values $X^2\!=\!175$ and $X^2\!=\!145$ correspond to the mean radii $\sqrt{175}\simeq 13.2$ and $\sqrt{145}\simeq 12$, both of which are relatively close to the core radius $R\!=\!10.26$ inside the surfaces.

\begin{figure}[ht!]
\centering
\includegraphics[width=13.8cm]{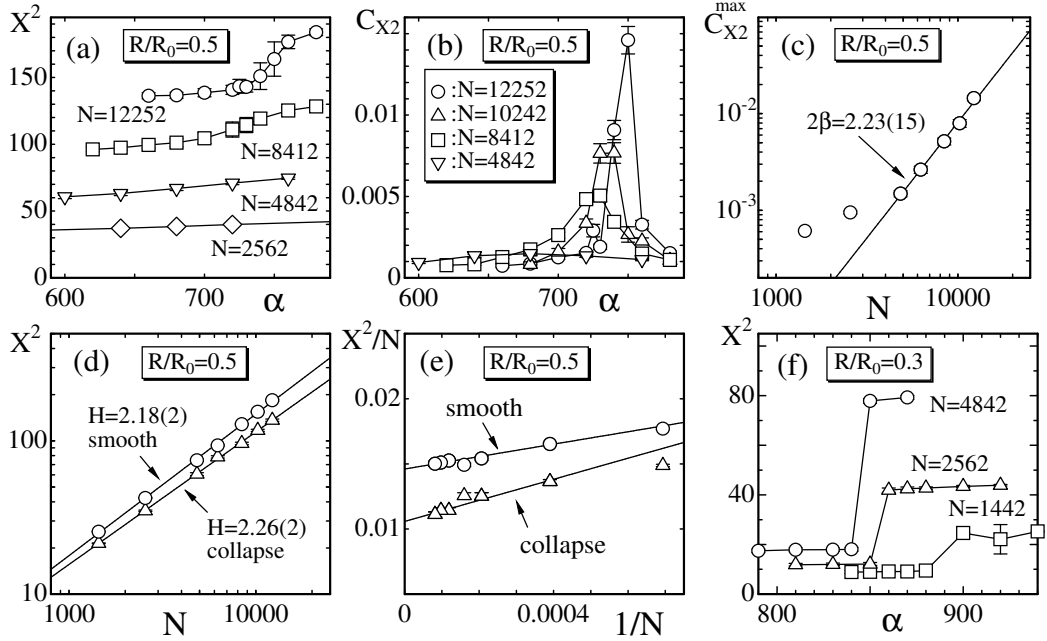}
\caption{%
(a) The mean square size {$X^2$} vs. {$\alpha$} at $R/R_0\!=\!50\%$, (b) the variance {$C_{X^2}$} vs. {$\alpha$}, (c) a log-log plot of {$C_{X^2}^{\rm max}$} vs. {$N$}, (d) {$X^2$} vs. {$N$} in the smooth and the collapsed phases at the transition point, (e) {$X^2/N$} vs. {$1/N$}, and (f) {$X^2$} vs. {$\alpha$} at {$R/R_0\!=\!30\%$}. The solid lines connecting the symbols in (a), (b), and (f) are drawn as a guide to the eyes. 
}
\label{fig-3}
\end{figure}
The mean square size $X^2$ in Eq. (\ref{X2}) is plotted against $\alpha$ in Fig. \ref{fig-3}(a), where $R/R_0\!=\!50\%$.  Figure {\ref{fig-3}(b)} is the variance of {$X^2$} defined by
\begin{equation}
\label{CX2}
 C_{X^2}=\frac{1}{N} \langle \left(X^2-\langle X^2\rangle \right)^2 \rangle.
\end{equation}
The peak values {$C_{X^2}^{\rm max}$} vs. {$N$} are shown in Fig. {\ref{fig-3}(c)} in a log-log scale. The straight line in Fig. {\ref{fig-3}(c)} is drawn by fitting the largest five data to  
\begin{equation}
\label{CX2-scaling}
C_{X^2}^{\rm max}\sim N^{2\beta}, \quad 2\beta=2.23\pm 0.15.
\end{equation}
Thus, we have {$\sqrt{C_{X^2}^{\rm max}}\sim N^{\beta}$} with {$\beta\!=\!1.17\pm 0.08$}, and therefore, we find that the transition is of first-order because of the finite-size scaling theory {\cite{PRIVMAN-WS-1989,BINDER-RPP-1997}}. 

The Hausdorff dimension {$H$} is defined by
\begin{equation}
\label{H}
 X^2\simeq N^{\frac{2}{H}} \quad (N\to \infty).
\end{equation}
To obtain {$H$} at the transition point at {$R/R_0\!=\!50\%$}, we show {$X^{2,{\rm smo}}$} and {$X^{2,{\rm col}}$} in Fig. {\ref{fig-3}(d)}, where {$X^{2,{\rm smo}}$} and {$X^{2,{\rm col}}$} denote {$X^2$} in the smooth phase and in the collapsed phase, respectively. 
Since {$X^{2,{\rm smo}}$} and {$X^{2,{\rm col}}$} are hardly obtained at the transition point, we use the largest (smallest) {$X^2$} at every {$N$} in Fig. {\ref{fig-3}(a)} as {$X^{2,{\rm smo}}$} ({$X^{2,{\rm col}}$}). By fitting the data excluding {$X^2(N\!=\!6252)$} to Eq. ({\ref{H}}), we have 
\begin{equation}
\label{Hausdorff}
H=2.18\pm 0.02\; ({\rm smooth}), \qquad H=2.26\pm 0.02\; ({\rm collapsed}).
\end{equation}
The value of {$H$} in the collapsed phase is very close to {$H$} in the smooth phase as expected.

Although $S_1/N\not= 1.5$ due to the potential {$U(R)$}, {$X^{2,{\rm col}}/N$} is expected to be a non-zero finite constant in the limit of {$N\!\to\!\infty$}. In fact, {$X^{2,{\rm col}}$} is larger than the core radius squares {$R^2$} and smaller than $X^{2,{\rm smo}}$. Recalling that both {$R^2/N$} and $X^{2,{\rm smo}}/N$ are independent of $N$ at  {$N\!\to\!\infty$} and that {$R^2<X^{2,{\rm smo}}$, we expect that {$X^{2,{\rm col}}$} is independent of $N$ at {$N\!\to\!\infty$}. Figure {\ref{fig-3}(e)} shows {$X^{2,{\rm col}}/N$} and {$X^{2,{\rm smo}}/N$} against {$1/N$}, where the straight lines are drawn by the linear fitting of the data excluding the data of {$N\!=\!1442$} and {$N\!=\!6252$}. }     

We see from Fig. {\ref{fig-3}(f)} that {$X^2$} at {$R/R_0\!=\!30\%$} discontinuously changes even on the small surfaces. The transition point $\alpha_c(N)$ is not always identical with $\alpha$ where some physical quantities discontinuously change. A discontinuity seen $X^2$ in Fig. \ref{fig-3}(f) at $R/R_0\!=\!30\%$ only implies a possibility that the surface configuration is trapped in a potential minimum.  However, in this case the discontinuity clearly reflects a first-order nature of the transition. To the contrary, the transition point {$\alpha_c(N)$} is almost clear in Fig. {\ref{fig-3}}(a) at {$R/R_0\!=\!50\%$} because physical quantities vary continuously, such as {$X^2$ of the {$N\!=\!12225$} surface in Fig. {\ref{fig-3}}}(a) at {$R/R_0\!=\!50\%$}, although a discontinuous nature of the transition is unclear.   

\begin{figure}[ht!]
\centering
\includegraphics[width=10cm]{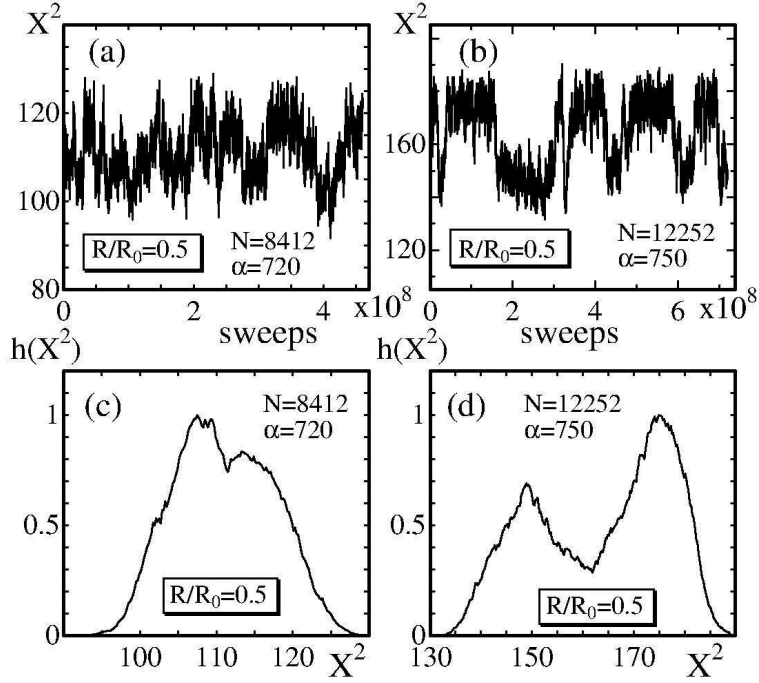}
\caption{Variations of $X^2$ vs. MCS of (a) the $N\!=\!8412$ surface at $\alpha\!=\!720$ and (b) the $N\!=\!12252$ surface at $\alpha\!=\!750$, (c) a normalized distribution $h(X^2)$ corresponding to the variation in (a), and (d) $h(X^2)$ corresponding to the variation in (b).  } 
\label{fig-4}
\end{figure}
To see the discontinuity in $X^2$ more clearly on the surfaces at $R/R_0\!=\!50\%$, we show variations of $X^2$ against MCS in Figs. \ref{fig-4}(a) and \ref{fig-4}(b). The discontinuity of $X^2$ is hardly seen in the variation of $X^2$ on the $N\!=\!8412$ surface, while it is clearly seen on the $N\!=\!12252$ surface. The corresponding normalized distribution $h(X^2)$ of $X^2$ is shown in  Figs. \ref{fig-4}(c) and \ref{fig-4}(d). A double peak structure can be seen in both of $h(X^2)$, and the peaks are clearly separated as $N$ increases. This implies a discontinuous transition. 

\begin{figure}[ht!]
\centering
\includegraphics[width=13.8cm]{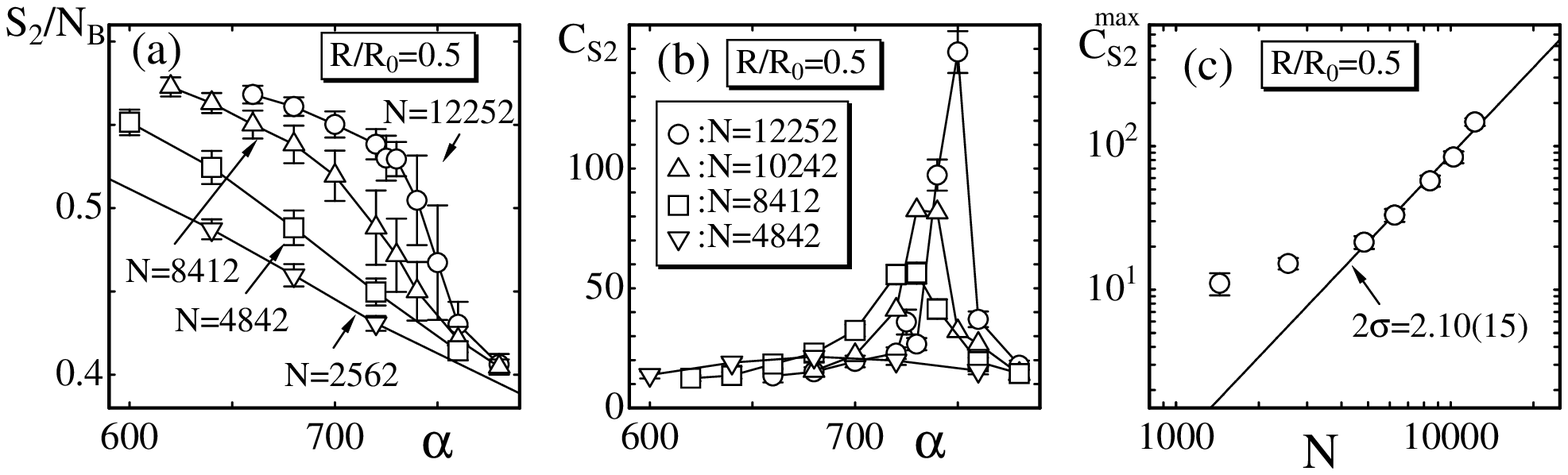}
\caption{(a) The bending energy $S_2/N_B$ vs. $\alpha$ at 
$R/R_0\!=\!50\%$ and , and (b) the variance {$C_{S_2}$} vs. {$\alpha$}, and (c) {$C_{S_2}^{\rm max}$} vs. {$N$} in a log-log scale. The symbol $S_2$ is used to distinguish the extrinsic curvature energy from the intrinsic curvature one $S_2^{\rm int}$.  } 
\label{fig-5}
\end{figure}
The bending energy $S_2$ is defined by
\begin{equation} 
\label{bending-energy}
 S_2 = \sum_{ij} \left(1-{\bf n}_i\cdot{\bf n}_j\right),
\end{equation} 
where ${\bf n}_i$ is a unit normal vector of the triangle $i$, and $\sum_{ij}$ is the sum over all pairs of nearest neighbor triangles $i$ and $j$. The surface shape is reflected in this $S_2$, which is not included in the Hamiltonian. $S_2/N_B$ vs. $b$ is shown in Fig. \ref{fig-5}(a), where $N_B$ is the total number of bonds.  The variance of {$S_2$} is defined by
\begin{equation}
\label{specific_heat}
C_{S_2}=\frac{1}{N} \langle \left(S_2-\langle S_2\rangle \right)^2 \rangle,
\end{equation}
and it is plotted against {$\alpha$} in Fig. {\ref{fig-5}}(b). The peaks {$C_{S_2}^{\rm max}$} vs. {$N$} are plotted in Fig. {\ref{fig-5}}(c) in a log-log scale. The straight line is drawn by fitting the largest five data to the relation
\begin{equation}
\label{CS2-scaling}
C_{S_2}^{\rm max}\sim N^{2\sigma}, \quad 2\sigma = 2.10\pm 0.15. 
\end{equation}
Thus, we have {$\sigma\!=\!1.05\pm0.08$}, where {$\sqrt{C_{S_2}^{\rm max}}\sim N^{\sigma}$}. This result implies a discontinuous transition of surface fluctuations, and the result is consistent with that of Eq. ({\ref{CX2-scaling}}).

\begin{figure}[ht!]
\centering
\includegraphics[width=10cm]{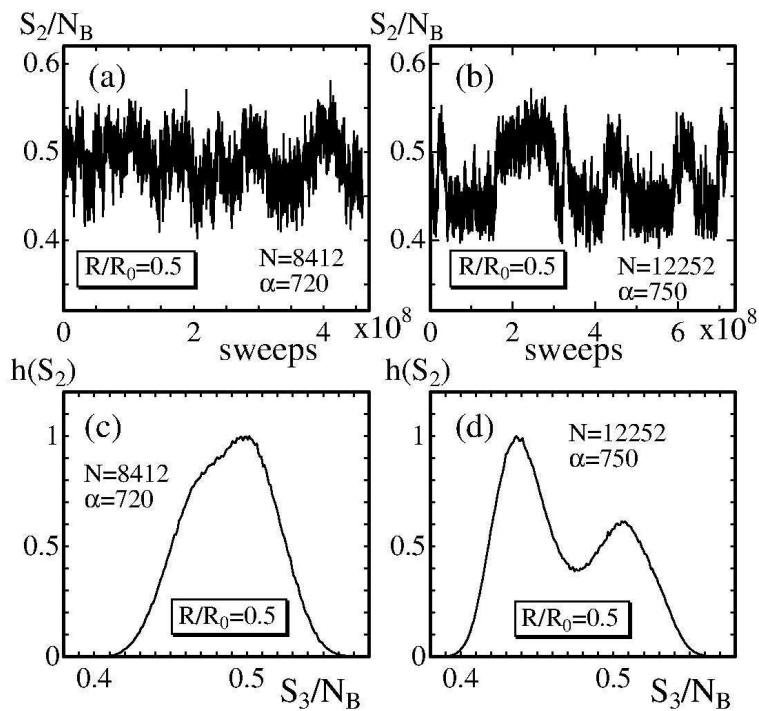}
\caption{The variation of $S_2/N_B$ vs. MCS of (a) the $N\!=\!8412$ surface at $\alpha\!=\!720$ and (b) the $N\!=\!12252$ surface at $\alpha\!=\!750$, (c) the normalized distribution $h(S_2)$ corresponding to (a), and (d) the normalized distribution $h(S_2)$ corresponding to (b). } 
\label{fig-6}
\end{figure}
The variation of $S_2/N_B$ against MCS is plotted in Figs. \ref{fig-6}(a) and \ref{fig-6}(b), where $S_2/N_B$ reflects the shape fluctuations. A discontinuity of $S_2/N_B$ is clearly seen in the variation of Fig. \ref{fig-6}(b). Figures \ref{fig-6}(c) and \ref{fig-6}(d) show the corresponding normalized histograms $h(S_2)$. A double peak structure is clearly seen in $h(S_2)$ of Fig. \ref{fig-6}(d), while it is not in Fig. \ref{fig-6}(c). This also implies that the transition is of first order, because the double peak becomes apparent with increasing $N$. 

\begin{figure}[ht!]
\centering
\includegraphics[width=10cm]{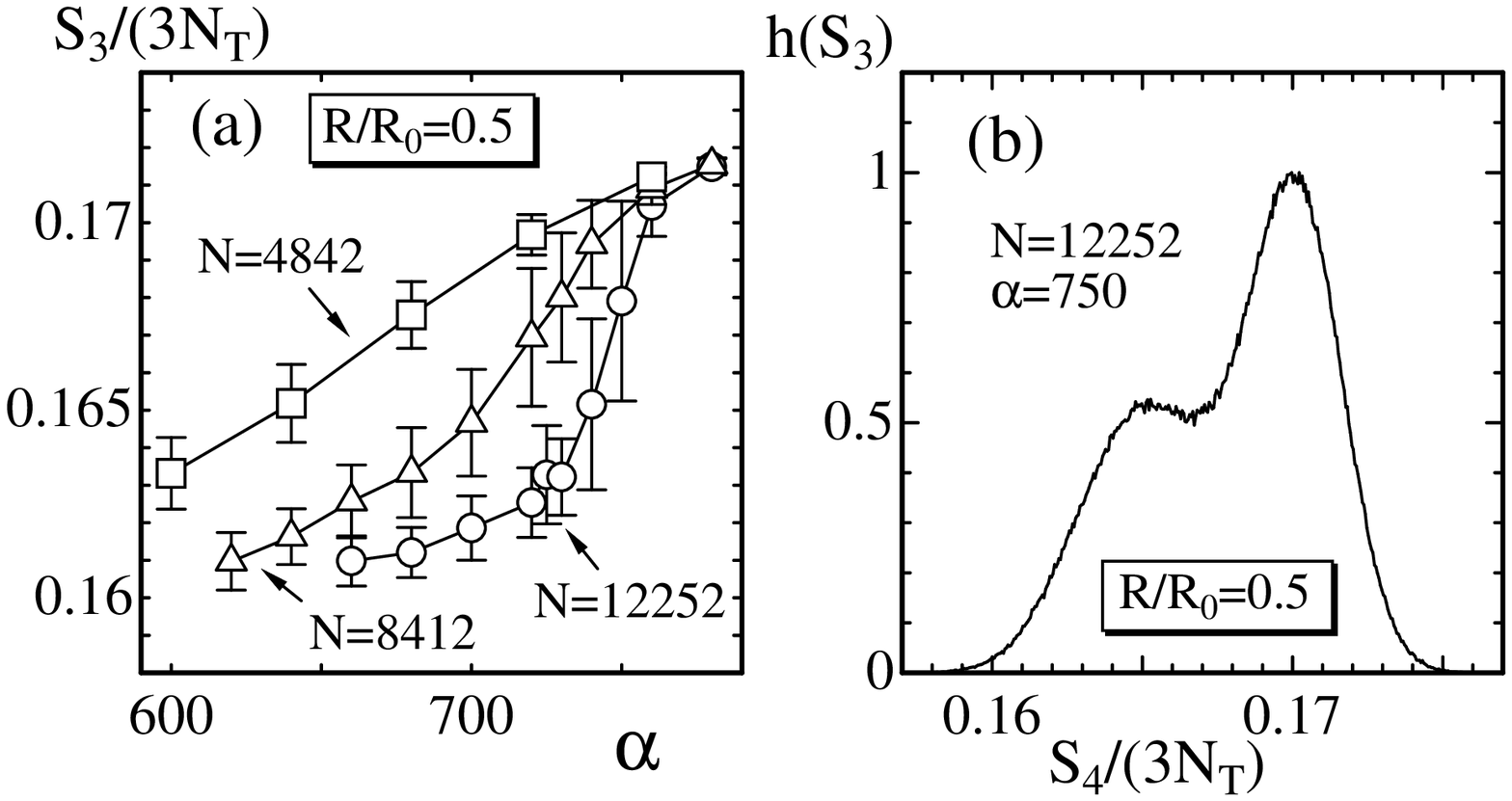}
\caption{(a) The in-plane shear energy $S_3/(3N_T)$ vs. $\alpha$ at $R/R_0\!=\!50\%$, and (b) the normalized histogram $h(S_3)$ of $S_3/(3N_T)$ on the $N\!=\!12252$ surface at $\alpha\!=\!750$. } 
\label{fig-7}
\end{figure}
The in-plane shear energy $S_3$ is defined by
\begin{equation} 
\label{shear-energy}
 S_3 = \sum_{i=1}^{3N_T} \left[1-\cos \left(\theta_i-\frac{\pi}{3}\right) \right],
\end{equation} 
where $\theta_i$ is an internal angle $i$ of a triangle, and $N_T(=2N\!-\!4)$ is the total number of triangles. $S_3$ reflects an in-plane deformation of triangulated surfaces \cite{KOIB-EPJB-2008-2}, and is not included in the Hamiltonian. 

Figure \ref{fig-7}(a) shows $S_3/(3N_T)$ vs. $\alpha$ at $R/R_0\!=\!50\%$. The variation of $S_3/(3N_T)$ vs. $\alpha$ is also expected to be discontinuous at sufficiently large $N$ just like the quantities $S_2$ and $X^2$, which characterize the out of plane deformation. The normalized histogram $h(S_3)$ is shown in Fig. \ref{fig-7}(b), and we see a double peak structure in $h(S_3)$. This implies that the shape transformation transition can also be reflected in internal quantities of the surface. We note that the discontinuity is clearly seen in $S_3/(3N_T)$ at least when $R/R_0\!\leq\!30\%$. 

\begin{figure}[ht!]
\centering
\includegraphics[width=13.8cm]{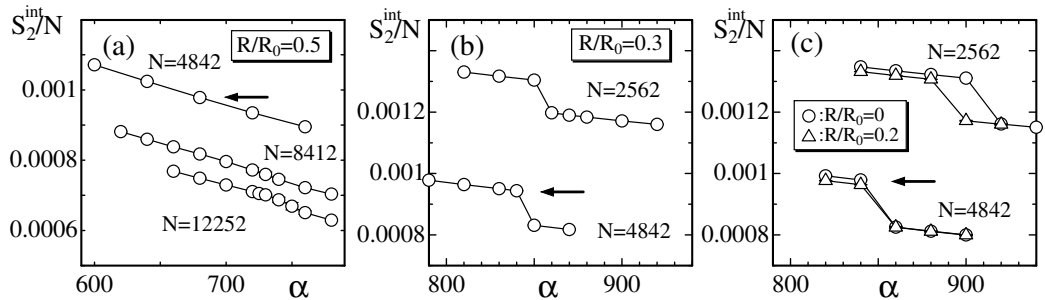}
\caption{The intrinsic curvature energy $S_2^{\rm int}/N$ vs. $\alpha$ at (a) $R/R_0\!=\!50\%$,  (b) $R/R_0\!=\!30\%$ (c) $R/R_0\!=\!20\%, 0\%$. The left arrows drawn at the $N\!=\!4842$ data indicate that $S_2^{\rm int}/N$ in the collapsed phase is almost independent of the ratio $R/R_0$ at the transition region. } 
\label{fig-8}
\end{figure}
The intrinsic curvature energy $S_2^{\rm int}/N$ is shown in Figs. \ref{fig-8}(a)--\ref{fig-8}(c). As mentioned above, $S_2^{\rm int}/N$ does not always have a clear jump at the transition under  $R/R_0\!=\!50\%$. The reason why $S_2^{\rm int}/N$ appears to vary continuously against $\alpha$  under  {$R/R_0\!=\!50\%$} is because the transition occurs many times during the simulations as we see in the variations of $X^2$ in Fig. \ref{fig-4}(b) and $S_2/N$ in Fig. \ref{fig-6}(b).

We should remark that $S_2^{\rm int}/N$ ($N\!=\!4842$) at $R/R_0\!=\!50\%$ in Fig. \ref{fig-8}(a) is almost comparable to $S_2^{\rm int}/N$ ($N\!=\!4842$) at $R/R_0\!=\!30\%$, $R/R_0\!=\!20\%$,  and $R/R_0\!=\!0\%$ in Figs. \ref{fig-8}(b)--\ref{fig-8}(c) in the collapsed phase close to the transition point. These values are pointed out by the left arrows in the figures. This observation allows us to consider that the collapsed phase is degenerate. The collapsed states belong to a single value of $S_2^{\rm int}/N$ or to a narrow and continuous range of $S_2^{\rm int}/N$. This degeneracy comes from the fact that $S_2^{\rm int}$ is insensitive to the surface shape as we have discussed in Section \ref{model}. 

\begin{figure}[h!]
\centering
\includegraphics[width=10cm]{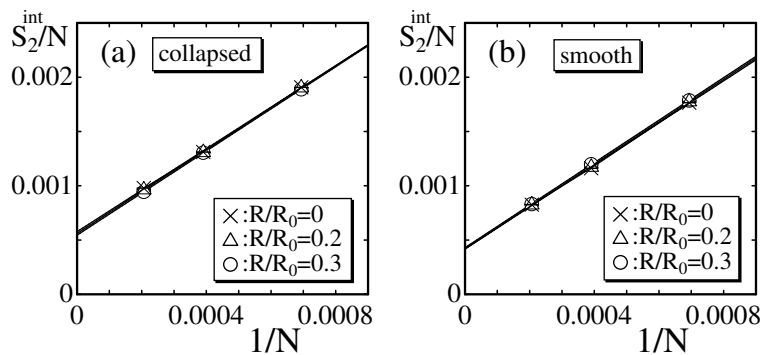}
\caption{ {$S_2^{\rm int}/N$} vs. {$1/N$} at (a) the collapsed phase and (b) the smooth phase just below and above the transition points. The data are obtained on the surfaces of {$N\!=\!4842$}, {$N\!=\!2562$}, and {$N\!=\!1442$}. The straight lines are drawn by the linear fitting of the data. In the limit of {$N\!\to\!\infty$}, {$S_2^{\rm int}/N$} is independent of $R/R_0(\leq \!0.3)$ in both of the collapsed phase and in the smooth phase at the transition point.} 
\label{fig-9}
\end{figure}
To see this degeneracy more clearly, we plot {$S_2^{\rm int}/N$} vs. {$1/N$} in Fig. {\ref{fig-9}}(a), where $N\!=\!4842$, $N\!=\!2562$, $N\!=\!1442$. The data {$S_2^{\rm int}/N$}, including those just indicated by the left arrows in Figs. {\ref{fig-8}}(b)--{\ref{fig-8}}(c),  are obtained at the transition point. We see no dependence of {$S_2^{\rm int}/N$} on the conditions {$R/R_0\!=\!0\%\sim 30\%$} and no difference in {$S_2^{\rm int}/N$} even in the limit of {$N\!\to\!\infty$}. In the smooth phase, on the other hand, such a degeneracy is exactly expected in {$S_2^{\rm int}/N$}, because the smooth surface is not influenced by the core inside it. Figure {\ref{fig-9}}(b) shows {$S_2^{\rm int}/N$} in the smooth phase. We should note that the size-effect is not expected in {$S_2^{\rm int}/N$}. In fact, the size-effect can be seen only in the variances shown in Figs. {\ref{fig-3}}(c) and {\ref{fig-5}}(c).

\begin{figure}[h!]
\centering
\includegraphics[width=13.8cm]{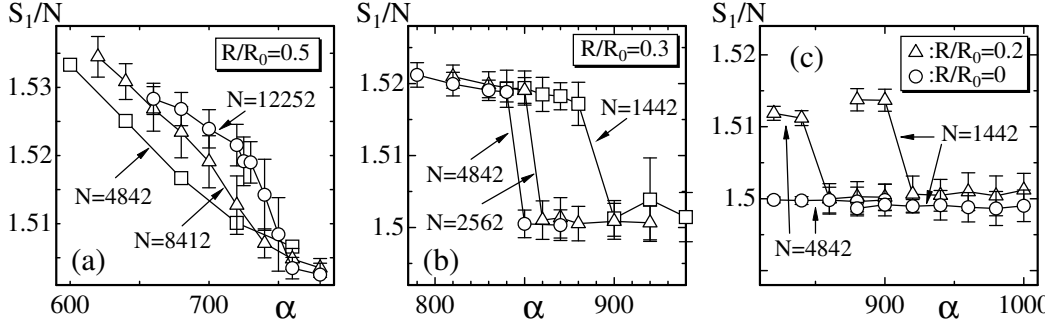}
\caption{The Gaussian bond potential $S_1/N$ vs. $\alpha$ at (a) $R/R_0\!=\!50\%$,  (b) $R/R_0\!=\!30\%$,  (c) $R/R_0\!=\!20\%, 0\%$.} 
\label{fig-10}
\end{figure}
Finally, we show $S_1/N$ vs. $\alpha$ in Figs. \ref{fig-10}(a) -- \ref{fig-10}(c). As mentioned in Section \ref{model}, the relation $S_1/N\!=\!3/2$ can be violated due to the potential $U(R)$. We see that $S_1/N$ discontinuously changes against $\alpha$ if $R/R_0\!\geq\!20\%$ at least, while the relation $S_1/N\!=\!3/2$ holds at $R/R_0\!=\!0\%$ and in the smooth phase at {$R/R_0\!\leq\!30\%$}. The variations of $S_1/N$ in Fig. \ref{fig-10}(a) look very similar to those of $S_2/N_B$ in Fig. \ref{fig-5}(a). If the gap in $S_1/N$ shown in Figs. {\ref{fig-10}}(b) and {\ref{fig-10}}(c) is very large, it can influence the surface size such as $X^2$. However, the gap in $S_1/N$ is very small compared to the value of $S_1/N$ itself, and consequently, the influence of the gap in $S_1/N$ on $X^2$, or on the surface size, is negligible. Nevertheless, the discontinuous change in $S_1/N$ indicates that the surface area discontinuously changes. We should note that the gap in {$S_1/N$} is seen on such small surface of {$N\!=\!1442$} at {$R/R_0\!=\!20\%$}. This implies that the core is effective as the potential {$U(R)$} in Eq. (\ref{Disc-Eneg_1}) at least under {$R/R_0\!\geq\!20\%$} on the surfaces of {$N\!\geq\!1442$}. Note also that the core slightly influences {$S_1/N$} even in the smooth phase; in fact, $S_1/N$ at {$R/R_0\!=\!20\%$} in the smooth phase is slightly different from the one at {$R/R_0\!=\!0\%$} on the {$N\!=\!1442$} surface. 

\section{Summary and Conclusion}\label{Conclusion}
In this paper, we numerically study a surface model of sphere topology focusing on whether or not the phase structure is influenced by a spherical core inside the surface. The model is an intrinsic curvature surface model, which is known to undergo a first-order transition between the smooth spherical phase and the collapsed phase on fixed-connectivity surfaces if it were not for the spherical core \cite{KOIB-EPJB-2004}. 

We find that the model undergoes a discontinuous transition even when the surface encloses the core. The strength of the transition depends on the core radius $R$; the transition strengthens (weakens) with decreasing (increasing) $R/R_0$, where $R_0$ is the radius of the equilibrium smooth surface at infinite curvature coefficient $\alpha\to \infty$. The transition is of first-order when $R/R_0\!\leq\!50\%$ at least. The results in this paper allow us to consider that the transition occurs at sufficiently large $R/R_0(<1)$ in the limit of $N\to \infty$. Moreover, we find also that the transition is non-trivially reflected in the Gaussian bond potential $S_1/N$ such that {$S_1/N$} discontinuously changes at the transition. The collapsing transition accompanies a discontinuous in-plane deformation such as the change of surface area.  

We should comment on the reason why the transition at one $R/R_0$ is identical to the transition at another $R/R_0$. This is because the collapsed states are degenerate, and the degeneracy can be understood as follows: the transition looks dependent on the ratio $R/R_0$ in the sense that the surface shape observed at one $R/R_0$ is different from the one at another $R/R_0$. In fact, the collapsed phase, separated from the smooth phase by the transition, looks strongly dependent on the ratio $R/R_0$; the collapsed surface is almost indistinguishable from the smooth surface at $R/R_0\!=\!50\%$, while the collapsed surface is completely different from the smooth surface in the limit of $R/R_0\!\to\!0$. However, the transitions are considered to be a single one, because all of those collapsed surfaces have almost the same value of the intrinsic curvature energy $S_2^{\rm int}/N$. Thus, we see that the multitude of the transitions comes from the fact that $S_2^{\rm int}$ is insensitive to the surface shape.

The results obtained in this paper imply that a large fluctuation of the density of membrane lipids is expected at the collapsing transition if the membrane encloses some artificial materials. 
In other words, the in-plane phase transition implies that the fluctuation of the lipid density, or of the distance between lipids, is divergent. Large fluctuations of these quantities are considered to be a signal of the phenomenon that the surface is almost broken. In fact, the obtained results indicate that a discontinuous change is observed in the area of surface, because $S_1/N$ is proportional to the area. Therefore, a discontinuous change of the density of vertices is also expected, because the surface area changes discontinuously while the total number of vertices $N$ is fixed.

Finally, we comment on the future study. It is very interesting to study a surface model with a variable metric $g$ on a two-dimensional surface $M$, from which a mapping $X$ is defined such that $X : M \mapsto  {\bf R}^3$ \cite{KOIB-NPB-2010}. In this paper, we show that the phase transition is seen in an "intrinsic curvature" model, in which the surface area is not allowed to change in the whole region of the curvature coefficient $\alpha$ whenever the surface encloses no core. To the contrary, in such an "extrinsic curvature" model with a variable $g$, the surface area of $X(M)$ is allowed to change at the collapsing transition even when the surface encloses no core. Therefore, the discontinuous change of $S_1/N$ in the extrinsic curvature model with variable $g$ is expected if the surface encloses a spherical core, and as a consequence, the in-plane deformation such as the change of surface area will be clarified in more detail.

\end{document}